\documentclass[twocolumn,amsmath,amssymb]{revtex4}
\usepackage{graphicx}
\usepackage{dcolumn}
\usepackage{bm}
\begin{document}
\title{Finite element method calculations of ZnO nanowires for nanogenerators}

\author{M. A. Schubert}
 \email{aschub@mpi-halle.mpg.de}
\author{S. Senz}
\author{M. Alexe}
\author{D. Hesse}
\author{U. G\"osele}
\affiliation{Max-Planck-Institut f\"ur Mikrostrukturphysik,
Weinberg 2, D-06120 Halle, Germany}

\date{\today}

\begin{abstract}
The bending of a nonconducting piezoelectric
ZnO nanowire is simulated by finite element method calculations.
The top part is bent by a lateral force, which could be applied by an 
atomic force microscope (AFM) tip. 
The generated electrical potential is $\pm 0.3~V$. 
This relatively high signal is, however, difficult to measure, due to
the low capacitance of the ZnO nanowire ($\sim 4\cdot 10^{-5}~pF$)
as compared to
the capacitance of most preamplifiers ($\sim 5~pF$).
A further problem arises from the semiconducting properties
of experimentally fabricated ZnO nanowires which causes
the disappearance of the voltage signal within picoseconds.
\end{abstract}

\maketitle

Recently 
nanogenerators for powering nanodevices were reported 
in which a ZnO \cite{nano-004}, GaN \cite{nano-006} 
or a CdS \cite{nano-005} nanowire 
converts mechanical
energy into electrical energy on bending the nanowire.
In these experiments the nanowires were bent
by an AFM tip.
Using finite element calculations Gao and Wang \cite{nano-002} 
calculated a piezoelectric potential
in the order of 0.3~V generated by bending a ZnO nanowire.
At first sight this high surface potential leaves the impression 
of an easy measurement and usability.
However, in practice there are several limiting factors
to be considered.
Firstly, the fabricated ZnO nanowires are
not perfect insulators, but rather n-doped semiconductors 
with a typical resistivity
of $1\ \Omega cm$.
The second problem is related to the actual measurement of the
signal generated by the bent nanowire.
In the present paper we will address shortly both problems 
and show
that the piezoelectrically generated signals by
ZnO nanowires can not easily be detected.
We discuss the possibility of measuring
the piezoelectric signals and the requirements for
energy harvesting.

\begin{figure}
\includegraphics[width=0.9\columnwidth]{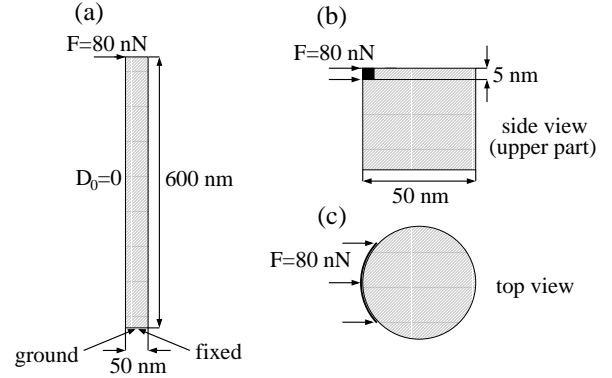}
\caption{\label{model} Schematic model of the ZnO nanowire.
(a) Side view of the whole nanowire.
Side view (b) and top view (c) of the upper part
with the applied force. }
\end{figure}

We calculated the bending of a ZnO nanowire 
using the program Comsol Multiphysics applying the 
finite element method (FEM) \cite{femlab}.
The nanowire was modelled as a perfect cylinder 
(Fig.~\ref{model}) of 600$\:$nm length and 25$\:$nm radius.
The bottom part of the nanowire was fixed and, 
in order to bend the wire, a lateral force 
of $F=80\;nN$ was applied
to the upper part.
For electrical boundary conditions we assume
a grounded surface element at the bottom and
no free charge.

The FEM program solves the following equations:
\begin{equation}
\sigma _{p}= c_{pq}\epsilon _{q}-e_{kp}E_{k}
\end{equation}
\begin{equation}
D_{i}=e_{iq}\epsilon _{q}+\kappa _{ik}E_{k}
\end{equation}
where $\sigma$ is  the stress tensor, $\epsilon$ the strain,
$c_{pq}$ the linear elastic constant,
$e_{iq}$ the piezoelectric coefficient, 
$\kappa _{ik}$ the dielectric constant,
$D$ the electric displacement and $E$ 
the electric field.
We used the elastic and piezoelectric constants
for ZnO from literature \cite{nano-002}.

An important detail for the FEM calculations is the geometry used 
to apply the force for bending the nanowire.
In literature the nanowire was bent by an AFM tip \cite{nano-004},
but in a simulation it is difficult to express this
realistically. The simplest method is to use a 
point force.
A force applied in one point is, however, not realistic and has 
the disadvantage of a non converging solution at this point. 
In a FEM calculation, the solution is always obtained on only a 
limited number of 
points, on the mesh. Thus, the result of the application of  
a non-physical point force depends on the mesh size. 
If a tip touches the nanowire, the contact area is at first very small,
but in the contact point the maximum loading of the tip is exceeded.
Therefore the contact area between the AFM tip and the nanowire increases.
To simulate a similar case we use a model of the nanowire which
consists of several cylinders stacked on top of
each other. 
The force was applied to a shell element of a small
cylinder (height of 5~nm) on top of the nanowire (Fig. \ref{model} (b), (c)~).

\begin{figure}
\includegraphics[width=\columnwidth]{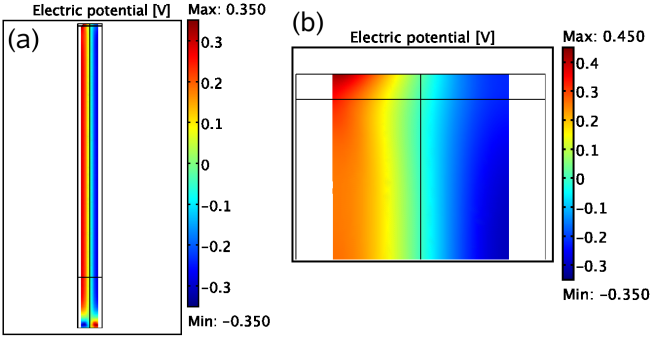}
\caption{\label{fem-1}FEM calculation of a ZnO nanowire 
bent by a lateral 
force of 80 nN. (a) the whole nanowire; 
(b) upper part of the nanowire. Here, the electric
potential on the shell element with the applied force
is slightly higher than in the rest of the nanowire. 
Note that the scale is different in both images
and that the nanowire is shown with its undistorted shape.}
\end{figure}

For an applied force of F = 80~nN the maximum 
bending (lateral deflection of the top) of the nanowire 
is 133~nm and the electric potential at 
a height of 300~nm is $\pm$0.3~V. 
Figure~\ref{fem-1} shows the results of the electric potential
for a bent ZnO nanowire,
the image on the left side shows the whole wire.
On the right side the top part of the nanowire is shown.
The electric potential is slightly increased on the left side of
the 5~nm high cylinder element on top of the nanowire,
at the place of the applied force. 

\begin{figure}
\includegraphics[width=0.58\columnwidth]{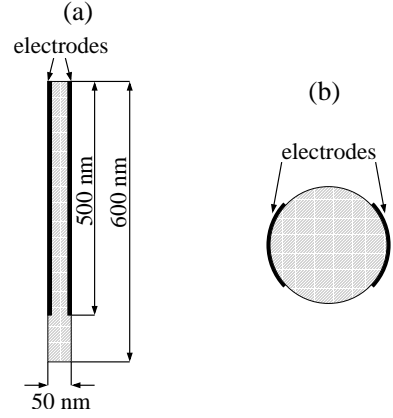}
\caption{\label{electrodes}ZnO nanowire with
electrodes on the sidewalls. (a) Side view and
(b) top view of the nanowire.}
\end{figure}

As the constitutive equations of the piezoelectric effect
show \cite{nano-007}, a piezoelectric sensor works in
converse piezoelectric mode and generates only charge
for a certain applied strain.
Therefore, a piezoelectric nanowire 
can not be regarded as a voltage source,
but rather as charge generator.
In order to estimate the charge that will
generate the calculated potential we need to
estimate the capacitance C of the nanowire.
Using the parallel plate capacitor approximation of
a nanowire with lateral electrodes as in Fig. \ref{electrodes},
C is given by:
\begin{equation}
C=\epsilon _{0}\epsilon _{r}\frac{A}{d}
\end{equation}
where $\epsilon _{0}=8.854\cdot 10^{-12}\ F\cdot m^{-1}$
is the vacuum permittivity and $\epsilon _{r}$ = 8.91 the
permittivity of the ZnO nanowire. For the area A of the capacitor 
we use a height of 500~nm and a width of 50~nm and for the
distance d the same value as for the width. For the
above dimension the capacitance of the nanowire (nw) is
$$
C_{nw}=3.9\cdot 10^{-17}\ F
$$
and the charge Q on the nanowire that will generate
the calculated potential should be
$$
Q_{nw}=C_{nw}\cdot U_{nw}=1.2\cdot 10^{-17}\ C.
$$
The above values are calculated assuming 
the most advantageous case when the
electrodes are deposited on the sidewalls
of the nanowire, 
and the whole generated charge is collected for the signal
generation. In real experiments
an AFM tip was used to bend the nanowire and also
to collect the signal \cite{nano-004}. 
In this case the capacitor area is
equal to the tip sample contact area
and it is several orders of magnitude smaller than above.
Therefore, to compare to real experiments we
should assume a contact area of around 200~nm$^{2}$. 
Thus, the effective 
capacitance will be reduced by a factor of 100 
to about $C_{tip\ contact}\approx 4\cdot 10^{-19}\  F$, likewise
the charge, that will be collected by the AFM tip,
which is now 
$Q_{tip\ contact}\approx 1.2\cdot 10^{-19}\ C$
($U_{tip\ contact}\approx U_{nw}$). 
The above rough estimation gives a signal 
equivalent to one electron that should be
collected by the AFM tip.
Using the FEM calculation we estimated the charge of a 
cylindric nanowire  with electrodes on the 
sidewalls as given in Fig. \ref{electrodes}
and for the AFM tip contact with the same area as for
the applied force.
The results calculated using FEM confirm our
rough estimation ($Q_{nw\ FEM}=1.7\cdot 10^{-17}~C$
and $Q_{tip\ contact\ FEM}\approx 1.7\cdot 10^{-19}~C$).

\begin{figure}
\includegraphics[width=0.8\columnwidth]{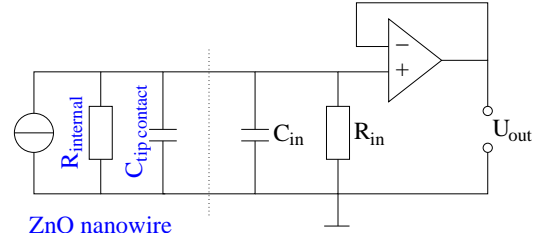}
\caption{\label{amplifier}Circuit with an amplifier 
for measuring the signal generated by a bent ZnO nanowire.
$R_{internal}=\infty$ in the ideal case of a nonconducting 
nanowire. ($R_{in}=100~M\Omega$, $C_{in}=5~pF$)}
\end{figure}

Even in the case of an ideal nonconducting ZnO nanowire, 
it is very difficult to transform the 0.3~V electric potential
in an output signal.
This potential is generated only by connecting an external 
circuit to this point (top of the nanowire). A typical
external circuit using an amplifier is given in
Fig.~\ref{amplifier}.
Most preamplifiers have an input capacitance of
$C_{in}\sim~5~pF$, but the capacitance of the nanowire using
an AFM tip as electrode is only about
$4\cdot 10^{-7}\ pF$.
Considering the most advantageous situation
of a direct coupling between the nanowire and the amplifier
input, from the charge conservation one can calculate
the voltage at the preamplifier input:
\begin{equation}
Q=C_{tip\ contact}\cdot U_{tip\ contact}=C_{in}\cdot U_{in}.
\end{equation}
This shows that in reality the calculated potential
of $0.3~V$ generates a signal of only
$20~nV$.
To measure such a low voltage represents quite a challenge.
The signal increases to $2~\mu V$ when the sidewalls 
of the ZnO nanowire are used as electrodes (as plotted in Fig. \ref{electrodes}).

Until now we discussed only an ideal non-conducting ZnO nanowire. 
However, in reality the fabricated ZnO nanowires are n-doped semiconductors
with a typical resistivity of 1~$\Omega$cm.
This value can be estimated from the I-V
characteristics given in literature \cite{nano-004, nano-001}.
The recent literature on ZnO nanowires confirms that
the typical resistivity values are in a range of
$10^{-2}~\Omega cm$ to $10~\Omega cm$.
The distribution of the potential as shown in Fig. \ref{fem-1}
suggests that the piezoelectrically generated charge is
distributed on the sidewalls and the resulting
electric field points perpendicular to the
nanowire axis. Therefore, the effective resistance
should be calculated accordingly:
\begin{equation}
R=\rho\cdot\frac{l}{A}
\label{resistivity}
\end{equation}
where l is the nanowire diameter and A the cross section of
the wire, in the approximation of a square wire.
This will give an effective resistance of
$20~k\Omega$ for $\rho = 1~\Omega cm$.

FEM calculation gives a
similar value of the resistance ($R=22.5~k\Omega$).
As we mentioned, the piezoelectric effect generates charge
and the time dependence of the generated voltage
can be calculated, using the classical textbook equation
for an RC circuit.
\begin{equation}
u(t)=U_{0}\cdot e^{-\frac{t}{R\cdot C}}
\end{equation}
where $u(t)$ is the time-dependent output voltage,
$U_{0}$ the initial voltage, t the time and 
$\tau=R\cdot C$ is the discharging time constant. After 
the time $\tau$ the output voltage decreases to 1/e of its 
initial value.
For the ZnO nanowire we obtain a discharging
time constant of $\tau=7.8\cdot 10^{-13}~s$
using the estimated value of the charge and the resistivity.
For the values calculated using the FEM we obtain:
$
\tau_{FEM}=1.3\cdot 10^{-12}~s
$.
This means simply that in only a few picoseconds the charge
generated by the piezoelectric effect will be cancelled
by internal losses.

In the experiments reported in the literature,
the nanowires were bent by an AFM tip with a limited scan speed.
For a scan speed of 100~$\mu$m/s the ideal nanowire is
completely charged after a time t$_{0}$=1.3~ms. 
The charge on the nanowire increases nearly linearly
during bending the wire.
Therefore we can calculate the charge current
$I_{c}=dq/dt$ for an ideal nanowire with electrodes
on the sidewalls ($Q_{nw}=1.2\cdot 10^{-17}~C$) during bending:
$$
I_{c}=\frac{Q_{nw}}{t_{0}}=9.0\cdot 10^{-15}~A.
$$

In the case of a real ZnO nanowire most of the charge current
flows through the resistor and does not charge the electrodes.
For a conductive nanowire ($R_{nw}=20~k\Omega$), the electric
potential is thus reduced from $0.3~V$ to:
$$
U_{real}=R_{nw}\cdot I_{c}=1.8\cdot 10^{-10}~V.
$$\\

In summary, using FEM calculations we have analyzed
a bent ZnO nanowire and we
obtained a piezoelectrically generated electric potential of 
about 0.3~V. 
This seems to be an easily readable signal, but in practice there
are important obstacles, even for an ideal ZnO nanowire.
The first is the very low capacitance
of the nanowire in case of the AFM tip contact 
($4\cdot 10^{-7}~pF$) compared to the input 
capacitance of a typical preamplifier ($5~pF$).
The corresponding charge generated by the nanowire is in
the order of the elementary charge.
According to charge conservation
the input voltage at the preamplifier is reduced from
$0.3~V$ to 
about $20~nV$. 
In an ideal case with electrodes on the sidewalls of
the nanowire the input voltage at the preamplifier
is about $2~\mu V$.
Further problems arise for a real ZnO nanowires, 
because these wires are not perfect insulators,
but rather  
n-doped semiconductors with a typical resistivity of
less than \mbox{1~$\Omega cm$}. 
Due to this high conductivity
the charged nanowire is discharged 
within picoseconds. 
During bending the nanowire 
with an AFM tip (at a scan speed of $100~\mu m/s$) the electric
potential on the nanowire is thus reduced from $0.3~V$ to 
$U_{real}=1.8\cdot 10^{-10}~V$ as a result of a nonrelativistic
calculation.
This means that a real ZnO wire has probably no charge, 
during
and after bending with an AFM tip.
This makes it nearly impossible to measure a piezoelectric
signal of a real ZnO nanowire, in spite of reports
to the contrary in the literature \cite{nano-004, nano-005, nano-009, nano-010}.


\end{document}